\documentclass[12pt]{iopart}

\usepackage{epsfig}
\usepackage{iopams}

\newtheorem{theorem}{Theorem}

\newtheorem{lemma}{Lemma}

\newcommand{\ket}[1]{\left | #1 \right\rangle}
\newcommand{\bra}[1]{\left \langle #1 \right |}
\newcommand{\half}{\mbox{$\textstyle \frac{1}{2}$}}

\renewcommand{\epsilon}{\varepsilon}

\begin{document}

\title[Entanglement in mutually unbiased bases]{Entanglement in mutually unbiased bases}

\author{M Wie\'sniak$^{1}$\footnote{Present address: Institute of Theoretical Physics and Astrophysics, University of Gda\'nsk, 80-952 Gda\'nsk, Poland.},
T Paterek$^{2}$\footnote{Author to whom any correspondence should be addressed. Email: tomasz.paterek@nus.edu.sg},
and A Zeilinger$^{1,3}$}

\address{
  $^1$ Vienna Center for Quantum Science and Technology (VCQ), Faculty of Physics, University of Vienna, Boltzmanngasse 5, 1090 Vienna, Austria \\
  $^2$ Centre for Quantum Technologies, National University of Singapore, 3 Science Drive 2, 117543 Singapore, Singapore \\
  $^3$ Institute for Quantum Optics and Quantum Information (IQOQI), Austrian Academy of Sciences, Boltzmanngasse 3, 1090 Vienna, Austria}

\begin{abstract}
One of the essential features of quantum mechanics is that most pairs of observables cannot be measured simultaneously.
This phenomenon is most strongly manifested when observables are related to mutually unbiased bases.
In this paper, we shed some light on the connection between mutually unbiased bases and another essential feature of quantum mechanics, quantum entanglement.
It is shown that a complete set of mutually unbiased bases of a bipartite system contains a fixed amount of entanglement, independently of the choice of the set.
This has implications for entanglement distribution among the states of a complete set. In prime-squared dimensions we present an explicit experiment-friendly construction
of a complete set with a particularly simple entanglement distribution.
Finally, we describe basic properties of mutually unbiased bases composed only of product states.
The constructions are illustrated with explicit examples in low dimensions.
We believe that properties of entanglement in mutually unbiased bases might be one of the ingredients to be taken into account to settle the question of the existence of complete sets.
We also expect that they will be relevant to applications of bases in the experimental realization of quantum protocols in higher-dimensional Hilbert spaces.
\end{abstract}


\maketitle

\section{Introduction}

Quantum complementarity forbids the simultaneous knowledge of almost all pairs of observables.
This impossibility is drawn to the extreme in the case of observables described by operators whose eigenstates form mutually unbiased bases (MUBs).
Two bases are said to be unbiased if any vector from one basis has an overlap with all vectors from the other basis that is equal in modulo.
The definition for a bigger set of MUBs means that the unbiasedness property holds for all pairs of these bases.
Accordingly, if we can perfectly predict a measurement result of one such observable corresponding to an eigenstate in one of the bases, 
then the results of all other observables corresponding to all other basis vectors of all other bases in the set remain completely uncertain.
One typical example of a set of three MUBs is the eigenbases of spin-$\half$ projections onto three orthogonal directions:
a spin-$\half$ state along one axis leaves us totally uncertain about the results along the orthogonal axes.

A spin-$\half$ particle is a two-level quantum system, a qubit, and clearly admits three MUBs.
A $d$-level quantum system, a qudit with pure states described in $d$ dimensional Hilbert space, can have at most $d+1$ MUBs \cite{WF1989}, and such a set is referred to as the complete set of MUBs.
The first explicit construction of the complete sets of MUBs was presented by Ivanovi\'c for $d$ being a prime number \cite{IVANOVIC}.
Subsequently, Wootters and Fields constructed the complete sets for prime-power $d$ \cite{WF1989}.
Since then, many explicit constructions have been derived and they are collected in a recent review \cite{REVIEW}.
If $d$ is not a prime power, the number of MUBs remains unknown although it is considered unlikely that a complete set of MUBs exists in these cases.
For example, the works \cite{BH2007,BW2008,arX,RLE2011} describe failed numerical attempts to find a complete set of MUBs in dimension 6.
In addition to this fundamental question, MUBs find applications in quantum tomography \cite{WF1989}, quantum cryptography \cite{BRUSS,B-PP,MOHAMED}, 
the Mean King problem \cite{VAA1987,AE2001,ARAVIND2003,HHH2005}, and other tasks.

Here we study the properties of entanglement between subsystems of a global system with a composite (i.e. nonprime) dimension as well as entanglement distribution among the states of MUBs.
We show that the amount of entanglement, as measured as a function of the linear entropy of a subsystem, present in states of a complete set of MUBs of a composite dimension always must have a nonzero value that is independent of a chosen set.
In other words, entanglement is always present in such a complete set of MUBs and it is always the same 
independent of the choice of the complete set, being solely a function of dimensions of subsystems.
Moreover, for global dimensionality that is big enough, practically all MUBs of a complete set contain entanglement.
We then show an experiment-friendly procedure that creates complete sets of MUBs in all dimensions $d=p^2$, which are squares of a prime number.
This procedure uses only one entangling operation, which is repeatedly applied to states of product MUBs to give the complete set.
Remarkably, the generated set consists of either product states or maximally entangled states.
Finally, we discuss the properties of MUBs consisting of product states only.
We believe that understanding entanglement in MUBs can lead on the practical side to novel applications and on the conceptual side to an understanding of why complete sets of MUBs can (not) exist for nonprime-power $d$.

\section{Conservation of entanglement}

Consider a bipartite system composed of subsystems $A$ and $B$, i.e. its global dimension is $d = d_A d_B$.
Any (hypothetical) complete set of MUBs allows for efficient quantum tomography as it reveals complete information about an arbitrary quantum state of the system \cite{WF1989, LKB2003}.
Hence we intuitively expect that the average entanglement over all the states constituting the complete set of MUBs shall be fixed with respect to some measure, independent of the choice of the bases.

This intuition is made rigorous in this section.
The relevant measure of entanglement is a function of the linear entropy of a reduced density operator.
The idea of the proof is to use the property of a complete set of MUBs called a complex projective $2$-design \cite{BARNUM2002,KR2005},
which here means that the entanglement averaged over a complete set of MUBs
is the same as the entanglement averaged over all pure quantum states.
The latter is constant due to known results in statistical mechanics \cite{LUBKIN1978}.
The message of this section, namely that the amount of entanglement is the same independent of a choice of the complete set of MUBs,
may be well-known to scientists working with designs,
but our proof is elementary and has immediate consequences for the distribution of entanglement among the states of MUBs.

\subsection{Complete sets of mutually unbiased bases and designs}

A complete set of MUBs is composed of $d+1$ bases, each basis of $d$ orthonormal vectors.
We denote by $\ket{j_m}$ the $j$th state of the $m$th basis, 
where for convenience we enumerate the states and the bases as
$j=0,\dots,d-1$ and $m=0,\dots,d$.
To introduce the notion of a $2$-design, one studies polynomials $\mathcal{P}(i) \equiv \mathcal{P}(x_1,x_2,y_1^*,y_2^* | i)$,
which are biquadratic in variables $x_1,x_2$ and separately in variables $y_1^*,y_2^*$,
where $x_i,y_i$ are any coefficients of arbitrary state $|i\rangle$ with respect to a fixed (say, standard) basis
and $^*$ denotes complex conjugation.
Any complete set of MUBs is known to be a complex projective $2$-design \cite{BARNUM2002,KR2005} 
because the average of any $\mathcal{P}(j_m)$ over states $\ket{j_m}$ is the same as the average with the Haar measure over all pure states:
\begin{equation}
\langle \mathcal{P}(j_m) \rangle_{\mathrm{MUBs}} = \langle \mathcal{P}(i) \rangle_{\mathrm{Haar}}.
\label{MUB-HAAR}
\end{equation}

\subsection{The conservation law}

In order to utilize the design property of the complete set of MUBs in the studies of entanglement,
we characterize the latter by the purity of a reduced density operator, say $\rho_{A|j_m} = \mathrm{Tr}_B(\ket{j_m} \bra{j_m})$:
\begin{equation}
\mathcal{P}(j_m) \equiv \mathrm{Tr}(\rho_{A|j_m}^2).
\end{equation}
This quantity acquires its minimum of $\frac{1}{d_A}$ for maximally entangled states and its maximum of unity for unentangled product states.
By `maximally entangled states' we mean pure states with maximal possible entropy for the smaller of the subsystems.
Note that due to the properties of the Schmidt decomposition it does not matter which subsystem is taken into account.
Moreover, the assumptions behind Eq.~(\ref{MUB-HAAR}) are fulfilled and, since per definition $\langle \mathcal{P}(j_m) \rangle_{\mathrm{MUBs}} = \frac{1}{d(d+1)} \sum_{m=0}^{d} \sum_{j=0}^{d-1} \mathrm{Tr}(\rho_{A|j_m}^2)$,
using the design property we write
\begin{equation}
\mathcal{E} \equiv \sum_{m=0}^{d} \sum_{j=0}^{d-1} \mathrm{Tr}(\rho_{A|j_m}^2) = d(d+1) \langle \mathrm{Tr}(\rho_{A|i}^2) \rangle_{\mathrm{Haar}}.
\end{equation}
In the last step, we use the result by Lubkin \cite{LUBKIN1978},
who studied how close the average reduced density operator is to a completely mixed state
and found that
\begin{equation}
\langle \mathrm{Tr}(\rho_{A|i}^2) \rangle_{\mathrm{Haar}} = \frac{d_A + d_B}{d+1}.
\end{equation}
Therefore, the sum of entanglement over all the states of any complete set of MUBs is fixed and equal to
\begin{equation}
\mathcal{E} = d_A d_B (d_A + d_B).
\label{ENT_CONST}
\end{equation}
Note that the right-hand side is symmetric with respect to $d_A$ and $d_B$,
which reflects the fact that we can as well study subsystem $B$.

Eq.~(\ref{ENT_CONST}) has two immediate consequences.
The first is that the distribution of entanglement among
different states of a complete set of MUBs can be arbitrary as long as there is a proper amount of it.
For example, Eq.~(\ref{ENT_CONST}) allows a complete set of MUBs to be formed by product and maximally entangled states
as well as solely by partially entangled states.

The second conclusion is that we cannot have a complete set of MUBs built entirely of product states or entirely of maximally entangled states. 
\begin{lemma}
Assume that $d_A \le d_B$.
In a complete set of MUBs which contains $d_A+1$ product MUBs,
all other bases contain only maximally entangled states.
\label{PROD-ENT}
\end{lemma}
Proof.
The sum of $\mathcal{P}(j_m)$ over the states of product MUBs equals $d_A d_B(d_A + 1)$.
The only possibility to obtain the value of (\ref{ENT_CONST}) is when for all the remaining
$d_A^2 d_B (d_B-1)$ states, $\mathcal{P}(j_m)$ acquires its minimal value of $\frac{1}{d_A}$. $\Box$

\section{Complete sets of mutually unbiased bases in prime-squared dimension}

We showed that the complete set of MUBs may be chosen as consisting of product bases and bases containing only maximally entangled states.
Here we present a construction of the complete sets with this property in dimension $d = p^2$ , where $p$ is prime.
The complete set will be generated from product MUBs with repeated application of a \emph{single} entangling operation, in our case the control-phase operation.
This makes our construction experiment-friendly.
Explicit examples of MUBs generated by this method together with their factorization into product or maximally entangled bases are presented in the Appendices.

\subsection{Complete sets of mutually unbiased bases in prime dimensions}

Before we present the new construction, let us briefly recall some of the known ones to which we will refer later on.
If $d = p$ is a prime number a complete set of $p+1$ MUBs was first found by Ivanovi\'c \cite{IVANOVIC}.
It is convenient to enumerate the bases as $m=0,\dots,p$
with $m=p$ corresponding to a standard basis, 
i.e. the basis in which the vectors of all other MUBs will be expressed.
To simplify the notation and if no confusion arises, we will write the vectors of the standard basis without any index, i.e. $\ket{s} \equiv \ket{s_p}$ enumerates the states of the standard basis.
The other $p$ MUBs have the Fourier-Gauss structure,
\begin{equation}
\ket{j_m} = \frac{1}{\sqrt{p}} \sum_{s=0}^{p-1} \alpha_p^{j s + m s^2} \ket{s} \quad \textrm{for } \quad m=0,\dots,p-1, \quad \textrm{and } \quad p>2,
\label{MUB_P}
\end{equation}
where $\alpha_p = \exp(i 2 \pi / p)$ is the complex $p$th root of unity.
The only exception to this formula is the case of $p=2$
where one needs to refer to an imaginary unit $i$,
the fourth rather than the square root of unity.
For low dimensions, we present these bases explicitly in the Appendices.

In odd-prime dimensions, a standard basis and a single MUB are sufficient to generate the complete set of MUBs
with an application of a single unitary:
\begin{equation}
W = \textrm{diag}[1,\alpha_p,\alpha_p^4,\dots, \alpha_p^{(p-1)^2}],
\label{W}
\end{equation}
which has the standard basis as the eigenbasis
and permutes all other MUBs, i.e. $W \ket{j_m} = \ket{j_{m+1}}$
with addition modulo $p$.

Alternatively, one can construct complete sets of MUBs using Heisenberg-Weyl operators in  prime dimensions,
\begin{eqnarray}
X=\sum_{s=0}^{p-1}|s+1\rangle\langle s|,\quad 
Z=\sum_{s=0}^{p-1} \alpha_p^s |s\rangle\langle s|,
\end{eqnarray}
with addition inside the kets modulo $p$.
These operators span a unitary operator basis with respect to the trace scalar product as
\begin{equation}
\mathrm{Tr} \Big[ \left(X^aZ^b \right)^\dagger X^cZ^d \Big] = p\delta_{a,c}\delta_{b,d}.
\end{equation} 
According to the general result of Bandyopadhyay \emph{et al.} \cite{BANDYOPADHYAY}, 
if one can group elements of the unitary operator basis into disjoint subsets of $d$ commuting operators (unity being the only common element of these sets), 
the common eigenbases of the commuting operators within each set are mutually unbiased.
In the case of a system of a prime dimension, the groups of commuting operators can be chosen as powers of the operators $Z$, $X$, $X Z$, $XZ^2$, $\dots$, $XZ^{p-1}$.
Their eigenbases define a complete set of MUBs.
It turns out that this set of MUBs is identical to the set of Eq.~(\ref{MUB_P}) up to the indexing of bases and states within bases.

\begin{lemma} 
Bases (\ref{MUB_P}) are the eigenbases of the operators $X$, $X Z$, $\dots$, $XZ^{p-1}$.
\label{L_UNIT_MUB_P}
\end{lemma}
Proof. Choosing the standard basis as the eigenbasis of $Z$, the eigenbasis of $X$ is readily the Fourier basis, i.e. $\{\ket{j_0}\}$.
Next note that for $m=1,\dots,p-1$ we have $\ket{j_m} = \frac{1}{\sqrt{p}} \sum_{s=0}^{p-1} \alpha_p^{(j+m) s - 2m \xi_s} \ket{s}$ with $\xi_s = s + \dots + (p-1) = \frac{1}{2}(p-s)(p+s-1)$.
The proof that these are exactly the eigenstates of $X Z^{2m}$  is given in Ref. \cite{BANDYOPADHYAY}.
Since $p$ is prime, $2m$ runs through all the powers of $Z$. $\Box$

These two methods of generating complete sets of MUBs in prime dimensions can be generalized to prime-power dimensions.
However, these generalizations require a knowledge of elements of finite fields theory; see e.g. \cite{REVIEW}.
We now present our physically motivated construction of the complete set of MUBs in prime-squared dimensions $d=p^2$.

\subsection{Two qubits}

We begin with a statement relating the number of MUBs to the possibility of swapping the states of subsystems.
The statement itself holds for arbitrary dimension $d=p^2$,
but it can be directly used to produce a complete set of MUBs of only two qubits.

\begin{lemma}
Assume $d=p^2$ and there exists unitary $U$ that commutes with the swap operation $S$,
and such that vectors $\{ U \ket{a_k b_l} \}$, with $k \neq l$, form an MUB with respect to all product symmetric MUBs defined as $\{\ket{a_m b_m }\}$.
Then $\{ U \ket{a_l b_k} \}$ is MUB with respect to all the bases mentioned above.
\label{L_SWAP}
\end{lemma}
Proof. The commutativity of $U$ and $S$ and the Hermiticity of $S$ imply $U = S U S$.
The assumed unbiasedness is expressed as $|\langle a_m b_m | U | a_k' b_l' \rangle|^2 = \frac{1}{p^2}$ for all bases $m=0,...,p$ and all vectors $\ket{a_m b_m}$ and $\ket{a_k' b_l'}$.
The computation of the overlap
\begin{equation}
| \langle a_k b_l | U^{\dagger} U | a_l' b_k' \rangle|^2 = | \langle a_k | a_l' \rangle \langle b_l | b_k' \rangle|^2 = \frac{1}{p^2},
\end{equation}
reveals that the basis from the thesis is unbiased to $\{U \ket{a_k b_l}\}$.
The commutativity with the swap operation is used to prove its unbiasedness with respect to all product bases:
\begin{equation}
|\langle a_m b_m | U \ket{a_l' b_k'} |^2 = | \langle a_m b_m | S U S | a_l' b_k' \rangle|^2 = | \langle \alpha_m \beta_m | U | \alpha_k' \beta_l' \rangle|^2 = \frac{1}{p^2},
\end{equation}
where the last equality follows from the assumed unbiasedness
and we put $\alpha = b, \beta = a, \alpha' = b',\beta'=a'$. $\Box$

Note that the two bases $\{U \ket{a_k b_l}\}$ and $\{U \ket{a_l b_k}\}$
are simply related by the swap operation because $U \ket{a_l b_k} = S U S \ket{a_l b_k} = S U \ket{\alpha_k \beta_l}$ with $\alpha = b$ and $\beta = a$.

In case of $d=4$, this lemma allows us to generate the complete set of MUBs starting with product MUBs.
There are three MUBs in dimension $2$ and therefore we begin with the following three product MUBs in dimension $4$: $\{\ket{a_0 b_0}\}$, $\{\ket{a_1 b_1}\}$ and $\{\ket{a_2 b_2}\}$. 
Consider now application of the control-phase (control-$Z$) operation
\begin{equation}
\mathcal{P}_2 = \frac{1}{2}(I \otimes I + I \otimes \sigma_z + \sigma_z \otimes I - \sigma_z \otimes \sigma_z),
\label{CPHASE2}
\end{equation}
where $I$ denotes a single qubit identity operator and $\sigma_z=\left(\begin{array}{cc}1&0\\0&-1\end{array}\right)$.
We apply the control-phase operation on the two qubits prepared in states of the form $\ket{a_0 b_1}$.
The effect is best explained using Pauli operators.
For a single qubit, we choose, in accordance with Appendix A, the basis $m=0$ as the eigenbasis of $\sigma_x=\left(\begin{array}{cc}0&1\\1&0\end{array}\right)$ and basis $m=1$ as the eigenbasis of $\sigma_y=\left(\begin{array}{cc}0&-i\\i&0\end{array}\right)$.
Therefore, the basis $\{\ket{a_0 b_1}\}$ is the eigenbasis of the commuting operators $\sigma_x \otimes I$ and $I \otimes \sigma_y$, and their products.
The control-phase operation maps these operators onto
\begin{eqnarray}
\mathcal{P}_2 (\sigma_x \otimes I) \mathcal{P}_2 & = & \sigma_x \otimes \sigma_z, \\
\mathcal{P}_2 (I \otimes \sigma_y) \mathcal{P}_2 & = & \sigma_z \otimes \sigma_y.
\end{eqnarray}
The common eigenstates of these new operators are maximally entangled Bell states.
Moreover, such Bell basis is mutually unbiased with respect to all our product MUBs.
This can be verified directly or by using, e.g., the result of Bandyopadhyay \emph{et al.} \cite{BANDYOPADHYAY}.
We apply this theorem to tensor products of Pauli operators, eigenbases of which define our MUBs,
i.e. the three product MUBs $\{\ket{a_m b_m}\}$ are defined by sets of commuting operators 
$\{I \otimes I,\! \sigma_x \otimes I,\! I\otimes\sigma_x,\!  \sigma_x\otimes\sigma_x\}$,
$\{I \otimes I,\! \sigma_y\otimes I,\! I\otimes\sigma_y,\! \sigma_y\otimes\sigma_y\}$
and $\{I \otimes I,\! \sigma_z\otimes I,\! I\otimes\sigma_z,\!\sigma_z\otimes\sigma_z\}$, respectively,
whereas the Bell basis $\mathcal{P}_2 \ket{a_0 b_1}$ is defined by $\{I \otimes I,\! \sigma_x \otimes \sigma_z, \! \sigma_z \otimes\sigma_y,\!  \sigma_y \otimes\sigma_x\}$.
Each set of four is clearly a set of commuting operators,
and according to the mentioned theorem their eigenbases form MUBs.
Since the $\mathcal{P}_2$ operation is manifestly invariant under a swap of qubits,
according to Lemma \ref{L_SWAP} we obtain the following complete set of MUBs:
$\{\ket{a_0 b_0}\}, \{\ket{a_1 b_1}\}, \{\ket{a_2 b_2}\}, \{\mathcal{P}_2 \ket{a_0 b_1}\}, \{\mathcal{P}_2 \ket{a_1 b_0}\}$,
which is explicitly presented in Appendix C.
Note that for this dimension application of Lemma \ref{L_SWAP}
has the same effect as the result of Ref. \cite{IFDTHEND1}
stating for general dimension that if there is a set of $d$ 
MUBs, then there also exists a set of $d+1$ of them.

\subsection{Two qupits}

Now we move to a system of a global dimension $d=p^2$ with $p>2$.
Two systems, each of prime dimension $p$, admit altogether $p^2+1$ MUBs. 
We shall show that they all can be generated via the multiple application of a single entangling operation on product bases. 
For this purpose, we present a lemma which reduces the number of unbiasedness conditions one needs to check.

\begin{lemma}
For $p>2$ assume there exists unitary $U$ such that $|\bra{a_m b_m} U^n \ket{a_{0}' b_n'}|^2 = \frac{1}{p^2}$
for all $0\leq a,b,a',b'\leq p-1$, $n=1,...,p-1$ and $m=0,...,p$ and that $[U,W \otimes I] = [U,I \otimes W] = 0$, where $W$ is defined in Eq. (\ref{W}).
Then the bases $\{U^\nu \ket{a_{\mu} b_{\mu+\nu}}\}$, with $\mu,\nu=0,...,p-1$, together with the standard basis $\{\ket{a_p b_p}\}$ form a complete set of MUBs.
Addition of indices is modulo $p$.
\label{L_TO_C_PHASE}
\end{lemma}
Proof. Consider an overlap between states of two bases of the proclaimed form
\begin{equation}
\mathcal{M} \equiv |\bra{a_{\mu} b	_{\mu + \nu}} (U^{\nu})^{\dagger} U^{\nu'} \ket{a_{\mu'}' b_{\mu'+\nu'}'} |^2 = |\bra{a_{\mu } b_{\mu + \nu}} U^{\nu' - \nu} \ket{a_{\mu'}' b_{\mu'+\nu'}'} |^2.
\label{OVERLAP_2P}
\end{equation}
Since $U$ commutes with individual cycling unitary $W \otimes I$ and $I \otimes W$, it also commutes with their products. 
In particular, we have $U^{\nu' - \nu} =(W^{\mu'}\otimes W^{\mu'+\nu}) U^{\nu' - \nu} (W^{-\mu'} \otimes W^{-mu'-\nu})$.
We insert this expression into (\ref{OVERLAP_2P}) and since none of the bases there is the standard basis, the effect is to shift the indices of the local bases and get
\begin{equation} 
\mathcal{M} = |\bra{a_{\mu-\mu'} b_{\mu -\mu'}} U^{\nu' - \nu} \ket{a_0' b_{\nu'-\nu}'} |^2 = \frac{1}{p^2},
\end{equation}
where the last equality follows from our assumptions. 
Similarly, overlap with the standard basis equals $|\bra{a_{p} b_{p}} U^{\nu} \ket{a_{\mu}' b_{\mu+\nu}'} |^2 = \frac{1}{p^2}$
which follows from our assumptions after noting that the standard basis is not shifted by $W$, whereas the index of the other local bases we shift by $-\mu$. $\Box$

Now we prove that the control-phase operation
can be used to generate a complete set of MUBs in all prime-squared dimensions.
The control-phase reads
\begin{equation}
\label{controlphase}
\mathcal{P}_p= \frac{1}{p} \sum_{a,b=0}^{p-1} \alpha_p^{- ab} Z^a \otimes Z^b.
\end{equation}

\begin{theorem}
In every dimension $d=p^2$ with $p>2$, there exists an integer $\theta$ such that $\mathcal{P}_p^{\theta}$ satisfies requirements of Lemma \ref{L_TO_C_PHASE}.
\end{theorem}
Proof. First note that since both $\mathcal{P}_p$ and $W$ are diagonal in the standard basis,  $[\mathcal{P}_p^{\theta},W\otimes I]=[\mathcal{P}_p^{\theta},I\otimes W]=0$ is fullfilled for any $\theta$.

To prove that the bases $\{ \mathcal{P}_p^{\theta n} \ket{a_0 b_n} \}$ are unbiased to bases $\{ \ket{a_m b_m} \}$,
we refer once more to the results of Bandyopadhyay {\em et al.} \cite{BANDYOPADHYAY}.
They show that MUBs in prime dimensions $\{ \ket{j_{n}} \}$ may be chosen as eigenstates
of sets of commuting operators $X^{\beta} Z^{2 \beta n}$ with $\beta = 0,\dots,p-1$ (see also Lemma \ref{L_UNIT_MUB_P}).
The idea of the present proof is to show that operators defining bases $\{\ket{a_0 b_n}\}$ are transformed under the application of the control-phase into a new set of distinct operators
which are all different from the operators defining bases $\{\ket{a_m b_m}\}$.
Since commutation relations are preserved under unitary transformations, the results of \cite{BANDYOPADHYAY} imply that the new operators define MUBs with respect to $\{\ket{a_m b_m}\}$.

The control-phase acts symmetrically on both subsystems and we have up to a global phase:
\begin{equation}
\mathcal{P}_p^{\theta n} \left( X^{\alpha} \otimes X^\beta Z^{2 \beta n} \right) \mathcal{P}_p^{-\theta n} = X^{\alpha} Z^{\beta \theta n} \otimes X^\beta Z^{2 \beta n + \alpha \theta n }.
\label{TRANSFORMED}
\end{equation}
Since for different values of $\alpha$ and $\beta$ the initial operators $X^{\alpha} \otimes X^\beta Z^{2 \beta n}$ were orthogonal with respect to the trace scalar product,
the final operators are also orthogonal, i.e. we generated a set of trace-orthogonal operators which can be partitioned into proper commuting subsets.
We now have to ensure that the generated set does not contain any operators determining product MUBs $\{ \ket{a_m b_m} \}$.
Since in the product MUBs the bases of $A$ and $B$ are the same, their defining feature 
is that operators determining the basis of $A$ commute with the operators determining the basis of $B$.
We check whether this commutation condition is satisfied by the operators on the right-hand side of (\ref{TRANSFORMED}).
The operators of $A$, i.e. $X^{\alpha} Z^{\beta \theta n}$, commute with the operators of $B$, i.e. $X^\beta Z^{2 \beta n + \alpha \theta n}$, if and only if \cite{BANDYOPADHYAY}:
\begin{equation}
n (\alpha^2 \theta + 2 \alpha \beta - \beta^2 \theta) = 0 \textrm{ mod } p.
\end{equation}
We are interested only in positive $n$ and therefore ask whether the bracket is a multiple of a prime $p$.
In other words, we are looking for the solution of the quadratic equation in the prime field $\mathcal{F}_p$. 
It is well known that for $p>2$ such equations have solutions if and only if
there exists a field element
\begin{equation}
\Delta = 2 \beta \sqrt{1+\theta^2}.
\end{equation}
Therefore, we need to choose such a value of $\theta$ that $1+\theta^2$ has no square root in the prime field $\mathcal{F}_p$. 
Since for any element $x$ in the field $\mathcal{F}_p$ we have $x^2=(p-x)^2$, we have no more than $\frac{1+p}{2}$ elements with square roots. 
That is, there exist an element $x$ having a square root such that the next element, $1+x$, does not have a square root.
Hence there always exists $\theta$ such that $\sqrt{1+ \theta^2}\notin \mathcal{F}_d$. $\Box$

Generally speaking, there is no universal choice of $\theta$ that is independent of $p$.
We have neither found a function that for any given $p$ returns $\theta$ such that $1+\theta^2$ does not have a square root in the field,
and our construction generates the complete set of MUBs.
A good guess of a useful value of $\theta$ is often $1$. 
Out of the first $1000$ odd prime numbers, the construction with $\theta=1$ fails in $494$ cases, while out of the first $10000$ odd primes it fails in $4988$ cases.
The lowest numbers for which this value does not produce the complete set of MUBs are $7$, $17$ and $23$.

\subsection{Three qubits}

A similar construction using multiple application of only one entangling operation does not seem to exist for more than two subsystems of prime dimensionality. 
However, more operations can be used for the task.
Here we show that three entangling gates can be used to produce a complete set of nine MUBs for three qubits.

We start with the global standard basis $\{\ket{abc}\}$ and eight other bases that do not involve any local standard basis, 
i.e. $\ket{a_k b_l c_m}$ with $k,l,m=0,1$.
We next apply to the basis $\{\ket{a_k b_l c_m}\}$ operation
\begin{eqnarray}
&\mathcal{G}_{klm} = \frac{1}{2}\left( I \otimes I \otimes I + Z^k \otimes Z^l \otimes Z^m + Z^{1-k} \otimes Z^{1-l} \otimes Z^{1-m} - Z \otimes Z \otimes Z\right).&\nonumber\\
\end{eqnarray}
The resulting complete set of MUBs is given in Appendix F.

\subsection{Wocjan-Beth construction}

We would also like to mention the Wocjan-Beth construction \cite{WocjanBeth}, which is so far the only known construction that gives more MUBs in composite dimensions than there are for the smallest prime-power subsystem. The construction is designed for systems divisible into two identical subsystems. 

The method utilizes two kinds of vectors. The first kind is the so-called incident vectors, $V$. Exactly $d$ of their $d^2$ entries are equal to $1$; the rest is $0$. The task is to find families of $d$ such vectors that satisfy the following requirements: within each family every pair of vectors is orthogonal, and two vectors from two different families have the scalar product equal to one. For example, for $d=2$ there are only three families of incident vectors,
\begin{eqnarray}
\left\{\left(\begin{array}{c}1\\1\\0\\0\end{array}\right),\left(\begin{array}{c}0\\0\\1\\1\end{array}\right)\right\},&
\left\{\left(\begin{array}{c}1\\0\\1\\0\end{array}\right),\left(\begin{array}{c}0\\1\\0\\1\end{array}\right)\right\},&
\left\{\left(\begin{array}{c}1\\0\\0\\1\end{array}\right),\left(\begin{array}{c}0\\1\\1\\0\end{array}\right)\right\}.
\end{eqnarray}
There is a one-to-one correspondence between families of incident vectors and mutually orthogonal Latin squares of order $d$.

The other type of vectors is phase vectors, $h$, which have $d$ complex entries, each of modulo 1. The two kinds are combined through operation ``$\uparrow$''. $h\uparrow V$ shall be understood as $V$ with the first non-zero element multiplied by the first entry of $h$, the next by the second, etc. One needs $d$ orthogonal vectors $h$ and combines every phase vector with the incident vector using $\uparrow$. After normalization we get as many MUBs as the number of incident vector families we found.

We would like to mention that when we choose vectors $h$ proportional to the rows of the Fourier matrix and the first two incident families in the most natural way (similarly to the example), two bases generated in this way possess a product structure whereas all others are maximally entangled.
The present work suggests that it might be possible to extend this set with the `missing' product bases, which would make the Wocjan-Beth construction even more powerful.

\section{Product mutually unbiased bases}

Our last topic is limitations on the number of MUBs and their entanglement,
which follow from the fact that some bases are formed by product states.
We call such bases product MUBs.
First we present a straightforward bound on the maximal number of product MUBs;
next we discuss classes of product MUBs to show that in every dimension
one has two product MUBs such that there is no other product MUB with respect to them.
There could still be entangled MUBs and we give an example in which this entanglement does not help us to build a complete set of MUBs.

\subsection{Maximal number}

We begin by showing that the only way to construct product MUBs in composite dimension $d_A d_B$
is to build them from MUBs in dimensions $d_A$ and $d_B$ separately.

\begin{lemma}
Two product bases $\{\ket{ab}\}$ and $\{\ket{a' b'}\}$ in dimension $d_A d_B$
are mutually unbiased if and only if $\ket{a}$ is mutually unbiased to $\ket{a'}$ in dimension $d_A$
and $\ket{b}$ is mutually unbiased to $\ket{b'}$ in dimension $d_B$.
\label{L_PROD_MUB}
\end{lemma}
Proof. If local bases are mutually unbiased, then clearly their product bases are also mutually unbiased.
Conversely, assume the product bases are MUBs, i.e. $|\langle a | a' \rangle |^2 |\langle b | b' \rangle |^2 = \frac{1}{d_A d_B}$ for all $a,a',b,b'$.
Since the right-hand side is positive, neither of the scalar products of the left-hand side is zero.
In particular, this implies that keeping $a,a',b$ fixed we have for all values of $b'$ that $|\langle b | b' \rangle |^2 = 1/ d_A d_B |\langle a | a' \rangle |^2$.
Since the squared moduli are the quantum probabilities they sum up to 
$\sum_{b'} |\langle b | b' \rangle |^2 = 1$, which implies that $|\langle a | a' \rangle |^2 = \frac{1}{d_A}$ and hence $|\langle b | b' \rangle |^2 = \frac{1}{d_B}$.  $\Box$

The maximal number of product MUBs follows as a corollary.
In a general dimension $d = d_1 \dots d_n$
there are at most $\min_j \mathcal{M}_{j}$ product MUBs,
where $\mathcal{M}_{j}$ is the maximal number of MUBs in dimension $d_j$.
Note that the maximal number of product MUBs is also a corollary to Lemma \ref{PROD-ENT}.

\subsection{Direct and indirect bases}

Not every set of product bases can be of the cardinality described below Lemma \ref{L_PROD_MUB}.
The crucial distinction between the product bases is whether
their states can be distinguished with (i) local measurements only
or with (ii) additional classical communication \cite{PATER_PLA}.

The bases (i) are of form $\{\ket{a}\ket{b}\}$ having all states $\ket{a}$ orthogonal in the first subspace
and all states $\ket{b}$ orthogonal in the second subspace.
We shall call them \emph{direct} product bases because
the matrix $P_{\mathrm{direct}}$ having vectors $\ket{a}\ket{b}$ as columns
is a tensor product of matrices $A$ and $B$
having as columns vectors $\ket{a}$ and $\ket{b}$, respectively:
\begin{equation}
P_{\mathrm{direct}} = A \otimes B.
\end{equation}
An example of a direct product basis is the tensor product of standard bases.

Product bases (ii) can be written as $\ket{a}\ket{b(a)}$,
i.e. for every fixed vector $\ket{a}$ orthogonality of the product basis
requires states of the second subsystem $\ket{b(a)}$ to be orthogonal,
but importantly for different states $\ket{a}$ the orthogonal bases of the second subsystem may be different.
The measurement in such product basis requires classical communication:
after measuring the first subsystem the result needs to be fed-forward
to a device measuring second subsystem in order to adapt its setting to a suitable basis.
We shall call such bases \emph{indirect} product bases.
In matrix notation, the matrix of an indirect product basis $P_{\mathrm{indirect}}$ 
cannot be written as a tensor product of matrices of local bases,
but rather is of the form
\begin{equation}
P_{\mathrm{indirect}} = \sum_{a} \ket{a} \bra{a} \otimes B(a),
\end{equation}
where the columns of matrix $B(a)$ are vectors $\ket{b(a)}$.
An example of an indirect product basis of two qubits is
\begin{eqnarray}
\ket{0_0} \ket{0_0}, & \qquad & \ket{1_0} \ket{0_1}, \nonumber \\
\ket{0_0} \ket{1_0}, & \qquad & \ket{1_0} \ket{1_1}.
\label{FOURIER4}
\end{eqnarray}

\subsection{Blocking product mutually unbiased bases}

A set of product MUBs is \emph{blocked} if there exists no other mutually unbiased product basis with respect to this set.
Indirect product bases lead to the minimal blocked set of product MUBs,
and they have consequences for completeness of sets containing them.

\begin{lemma}
In every composite dimension, there is a blocked set of two product MUBs.
\label{L_UNEXT}
\end{lemma}
Proof. The first basis is a standard basis: direct product basis.
The second basis is an indirect product basis exhausting
all possible MUBs for at least one subsystem.
We order the subsystems such that $d_1 \ge d_2 \ge \dots \ge d_n$.
In dimension $d_2$, there are at most $d_2$ MUBs with respect to the local standard basis.
Since $d_1 \ge d_2$, for every orthogonal vector in dimension $d_1$
one can have different orthogonal bases in dimension $d_2$ which exhaust the whole set of local MUBs.
According to Lemma~\ref{L_PROD_MUB} there is no other product MUB. $\Box$

A compact explicit example of an indirect product basis
that together with the standard basis forms blocking product MUBs
can be given in dimension being a power of a prime $d = p^r$,
which is regarded as the dimension of a Hilbert space of a set of $r$ elementary $p$-level systems.
For every subsystem there are exactly $p$ MUBs with respect to the local standard basis
and so is the number of distinguishable local states.
Therefore, the basis $\{\ket{(j_1)_0} \ket{(j_2)_{j_1}} \ket{(j_3)_{j_1}} \dots \ket{(j_r)_{j_{r-1}}}\}$
exhausts all allowed MUBs for all but the first subsystem.
Here we denoted by $\ket{(j_n)_m}$ the state of the $n$th elementary subsystem in the $m$th MUB.

The indirect product bases can block extendibility not only of a set of product MUBs
but also of MUBs in general with no restriction to product bases.
For example, in dimension $4$ the set of three MUBs composed of the standard basis,
the indirect product basis of Eq.~(\ref{FOURIER4}) and the Fourier basis 
cannot be extended by any other MUB \cite{GRASSL_P,MUB_OLS}.

\section{Conclusions}

We were studying aspects of entanglement in states of MUBs in composite dimensions.
Independently of the way a global system is split into subsystems,
there is no complete set of MUBs that does not contain entanglement.
In contrast, practically all MUBs are entangled as the dimension of at least one of the subsystems grows to infinity.
The higher the dimension of the total system,
the smaller the ratio of the number of product MUBs to the cardinality of the complete set of MUBs.
This cardinality is proportional to the total dimension $d$,
whereas the largest number of product MUBs is of the order of the smallest prime-power factor of $d$.
Therefore, the ratio is the highest if $d = p^2$ is a square of a prime,
and even in this case the cardinality of the complete set is a square of the cardinality of the product MUBs
and the ratio vanishes in the limit $d \to \infty$.

We showed that entanglement of states of \emph{any} complete set of MUBs is fixed.
This has consequences for the distribution of entanglement among the states of a complete set
and might be a useful hint for a search of the complete sets or one of the ingredients to (dis)prove their existence.
This conservation law holds true independent of a division into subsystems
and therefore perhaps an argument could be found that there is a finite set of divisions under which the entanglement cannot simultaneously match the proper value.
Another route to follow is to begin with a set of states with a proper amount of entanglement
and apply local operations and classical communication in order to search for a complete set of MUBs.

We also considered practical implementations of complete sets of MUBs
and showed that for two subsystems, each with the same prime number of orthogonal states,
the complete set can be generated via the multiple application of a single entangling operation on product states.
The outcomes of this construction together with other examples of MUBs are explicitly presented in the Appendices for low dimensions (see also Ref. \cite{SMALL_MUB}).

Turning to possible experiments, we note that there are various avenues for implementing quantum states in higher dimensions.
For photons these include multiports and spatial-mode superpositions \cite{RECK,WEIHS,MULTIPORT,OBRIEN,LANGFORD,OBRIEN_MULTIPORT} 
or Hermite-Gauss and Laguerre-Gauss modes, most notably orbital angular momentum states \cite{OAM_NATURE,PADGETT_OAM, PADGETT_REVIEW,NL_OAM,NL_GROUP}.

\ack

We acknowledge discussions with Markus Grassl and Huangjun Zhu.
This research is supported by ERC Advanced Grant QIT4QAD, FWF SFB-grant F4007 of the Austrian Science Fund, and the National Research Foundation and Ministry of Education in Singapore.

\section*{Appendix}

We present here explicit examples of complete sets of MUBs and for composite dimensions we emphasize division into entangled and product states.
The notation used is explained on the example of a qubit ($d=2$).

\appendix

% DIMENSION 2

\section{d=2}

The symbol $\ket{j_m}$ denotes the $j$th vector of the $m$th MUB.
The standard basis is either denoted with subscript $d$ or has no subscript at all:
\begin{equation*}
B_2
=
\left(\begin{array}{cc}
1&0\\
0&1
\end{array}\right)
=
\left\{\begin{array}{l}
|0 \rangle \\
|1 \rangle
\end{array}\right\}
=
\left\{\begin{array}{l}
|0_2 \rangle \\
|1_2 \rangle
\end{array}\right\}.
\end{equation*}
Note that when we write a basis as a matrix, we can freely permute columns, since it only changes the order of the vectors in the basis.
\begin{equation*}
\label{dim2}
B_0 =
\frac{1}{\sqrt{2}}
\left(\begin{array}{ccc}
1&1\\
1&-1
\end{array}\right)
=
\left\{\begin{array}{l}
|0_0\rangle\\
|1_0\rangle
\end{array}\right\}, \quad 
B_1
=
\frac{1}{\sqrt{2}}
\left(\begin{array}{ccc}
1&1\\
i & -i
\end{array}
\right)
=
\left\{\begin{array}{l}
|0_1\rangle\\
|1_1\rangle
\end{array}\right\}.
\end{equation*}

% DIMENSION 3

\section{d=3}

\begin{equation*}
\label{dim3}
B_3 =\left(\begin{array}{ccc}1&0&0\\0&1&0\\0&0&1\end{array}\right)
=
\left\{ \begin{array}{l}|0 \rangle\\|1 \rangle \\ |2 \rangle \end{array}\right\},
\end{equation*}
\begin{equation*}
B_0=\frac{1}{\sqrt{3}}\left(\begin{array}{ccc}1&1&1\\1&\alpha_3&\alpha_3^2\\1&\alpha_3^2&\alpha_3\end{array}\right)
=
\left\{ \begin{array}{l}|0_0\rangle\\|1_0\rangle \\ |2_0\rangle \end{array}\right\},
\end{equation*}
\begin{equation*}
B_1= \frac{1}{\sqrt{3}}\left(\begin{array}{ccc}1&1&1\\\alpha_3&\alpha_3^2&1\\\alpha_3&1&\alpha_3^2\end{array}\right)
=
\left\{\begin{array}{l}|0_1\rangle \\|1_1\rangle \\ |2_1\rangle \end{array}\right\},
\end{equation*}
\begin{equation*}
B_2=\frac{1}{\sqrt{3}}\left(\begin{array}{ccc}1&1&1\\\alpha_3^2&1&\alpha_3\\\alpha_3^2&\alpha_3&1\end{array}\right)
=
\left\{\begin{array}{l}|0_2\rangle\\|1_2\rangle \\ |2_2\rangle \end{array}\right\},
\end{equation*}
where $\alpha_d = \exp{2 \pi / d}$ is the complex $d$th root of unity.

% DIMENSION 4

\section{d=4}

The bases of this Appendix present explicitly the result of construction described in section 3.2 of the main text.

\begin{equation*}
\label{dim4a}
B_4
=
\left(\begin{array}{cccc}
1&0&0&0\\
0&1&0&0\\
0&0&1&0\\
0&0&0&1
\end{array}
\right)
=
\left\{\begin{array}{c}
|0\rangle\\
|1 \rangle
\end{array}
\right\}
\otimes
\left\{\begin{array}{c}
|0\rangle\\
|1\rangle
\end{array}
\right\},
\end{equation*}
\begin{equation*}
B_0
=
\frac{1}{2}
\left(\begin{array}{cccc}
1&1&1&1\\
1&-1&1&-1\\
1&1&-1&-1\\
1&-1&-1&1
\end{array}
\right)
=
\left\{\begin{array}{c}
|0_0\rangle\\
|1_0\rangle
\end{array}
\right\}
\otimes
\left\{\begin{array}{c}
|0_0\rangle\\
|1_0\rangle
\end{array}
\right\},
\end{equation*}
\begin{equation*}
B_1
=
\frac{1}{2}
\left(
\begin{array}{cccc}
1&1&1&1\\
i&-i&i&-i\\
i&i&-i&-i\\
-1&1&1&-1
\end{array}\right)
=
\left\{\begin{array}{c}
|0_1\rangle\\
|1_1\rangle
\end{array}
\right\}
\otimes
\left\{\begin{array}{c}
|0_1\rangle\\
|1_1\rangle
\end{array}
\right\},
\end{equation*}
\begin{eqnarray*}
B_2
& = &
\frac{1}{2}
\left(
\begin{array}{cccc}
1&1&1&1\\
i&-i&i&-i\\
1&1&-1&-1\\
-i&i&i&-i
\end{array}\right)
=
\frac{1}{\sqrt{2}}
\left\{
\begin{array}{c}
|0_1 \rangle |0_0 \rangle + i |1_1 \rangle |1_0 \rangle \\
|0_1 \rangle |0_0 \rangle - i |1_1 \rangle |1_0 \rangle \\
|1_1 \rangle |0_0 \rangle + i |0_1 \rangle |1_0 \rangle \\
|1_1 \rangle |0_0 \rangle - i |0_1 \rangle |1_0 \rangle
\end{array}\right\}
\\
& = &
\mathcal{P}_2
\left[
\left\{\begin{array}{c}
|0_0\rangle\\
|1_0\rangle
\end{array}
\right\}
\otimes
\left\{\begin{array}{c}
|0_1\rangle\\
|1_1\rangle
\end{array}
\right\}
\right],
\end{eqnarray*}
\begin{eqnarray*}
B_3
&=&
\frac{1}{2}
\left(
\begin{array}{cccc}
1&1&1&1\\
1&-1&1&-1\\
i&i&-i&-i
\\-i&i&i&-i
\end{array}
\right)
=
\frac{1}{\sqrt{2}}
\left\{
\begin{array}{c}
|0_0 \rangle |0_1 \rangle + i |1_0 \rangle |1_1 \rangle \\
|0_0 \rangle |1_1 \rangle + i |1_0 \rangle |0_1 \rangle \\
|0_0 \rangle |0_1 \rangle - i |1_0 \rangle |1_1 \rangle \\
|0_0 \rangle |1_1 \rangle - i |1_0 \rangle |0_1 \rangle
\end{array}\right\}
\\
& = &
\mathcal{P}_2
\left[
\left\{\begin{array}{c}
|0_1 \rangle\\
|1_1 \rangle
\end{array}
\right\}
\otimes
\left\{\begin{array}{c}
|0_0\rangle\\
|1_0\rangle
\end{array}
\right\}
\right],
\end{eqnarray*}
where the kets refer to MUBs for the two-level system (Appendix A)
and $\mathcal{P}_2$ is the control-phase operation between two qubits
defined in Eq. (\ref{CPHASE2}) of the main text.

% DIMENSION 5

\section{d=5}

\begin{equation*}
\label{dim5}
B_5
=
\left(\begin{array}{ccccc}
1&0&0&0&0\\
0&1&0&0&0\\
0&0&1&0&0\\
0&0&0&1&0\\
0&0&0&0&1\end{array}\right)
=
\left\{\begin{array}{c}
|0\rangle\\
|1\rangle\\
|2\rangle\\
|3\rangle\\
|4\rangle
\end{array}\right\},
\end{equation*}
\begin{eqnarray*}
B_0
&=&
\frac{1}{\sqrt{5}}
\left(\begin{array}{ccccc}
1&1&1&1&1\\
1&\alpha_5&\alpha_5^2&\alpha_5^3&\alpha_5^4\\
1&\alpha_5^2&\alpha_5^4&\alpha_5&\alpha_5^3\\
1&\alpha_5^3&\alpha_5&\alpha_5^4&\alpha_5^2\\
1&\alpha_5^4&\alpha_5^3&\alpha_5^2&\alpha_5
\end{array}\right)
=
\left\{\begin{array}{c}
|0_0\rangle\\
|1_0\rangle\\
|2_0\rangle\\
|3_0\rangle\\
|4_0\rangle
\end{array}\right\},
\end{eqnarray*}
\begin{eqnarray*}
B_1
& = &
\frac{1}{\sqrt{5}}\left(\begin{array}{ccccc}
1&1&1&1&1\\
\alpha_5&\alpha_5^2&\alpha_5^3&\alpha_5^4&1\\
\alpha_5^4&\alpha_5&\alpha_5^3&1&\alpha_5^2\\
\alpha_5^4&\alpha_5^2&1&\alpha_5^3&\alpha_5\\
\alpha_5&1&\alpha_5^4&\alpha_5^3&\alpha_5^2
\end{array}\right)
=
\left\{\begin{array}{c}
|0_1\rangle\\
|1_1\rangle\\
|2_1\rangle\\
|3_1\rangle\\
|4_1\rangle
\end{array}\right\},
\end{eqnarray*}
\begin{eqnarray*}
B_2
&=&
\frac{1}{\sqrt{5}}\left(\begin{array}{ccccc}
1&1&1&1&1\\
\alpha_5^2&\alpha_5^3&\alpha_5^4&1&\alpha_5\\
\alpha_5^3&1&\alpha_5^2&\alpha_5^4&\alpha_5\\
\alpha_5^3&\alpha_5&\alpha_5^4&\alpha_5^2&1\\
\alpha_5^2&\alpha_5&1&\alpha_5^4&\alpha_5^3
\end{array}\right)
=
\left\{\begin{array}{c}
|0_2\rangle\\
|1_2\rangle\\
|2_2\rangle\\
|3_2\rangle\\
|4_2\rangle
\end{array}\right\},
\end{eqnarray*}
\begin{eqnarray*}
B_3
&=&
\frac{1}{\sqrt{5}}\left(\begin{array}{ccccc}
1&1&1&1&1\\
\alpha_5^3&\alpha_5^4&1&\alpha_5&\alpha_5^2\\
\alpha_5^2&\alpha_5^4&\alpha_5&\alpha_5^3&1\\
\alpha_5^2&1&\alpha_5^3&\alpha_5&\alpha_5^4\\
\alpha_5^3&\alpha_5^2&\alpha_5&1&\alpha_5^4
\end{array}\right)
=
\left\{\begin{array}{c}
|0_3\rangle\\
|1_3\rangle\\
|2_3\rangle\\
|3_3\rangle\\
|4_3\rangle
\end{array}\right\},
\end{eqnarray*}
\begin{eqnarray*}
B_4&=&\frac{1}{\sqrt{5}}\left(\begin{array}{ccccc}
1&1&1&1&1\\
\alpha_5^4&1&\alpha_5&\alpha_5^2&\alpha_5^3\\
\alpha_5&\alpha_5^3&1&\alpha_5^2&\alpha_5^4\\
\alpha_5&\alpha_5^4&\alpha_5^2&1&\alpha_5^3\\
\alpha_5^4&\alpha_5^3&\alpha_5^2&\alpha_5&1
\end{array}\right)
=
\left\{\begin{array}{c}
|0_4\rangle\\
|1_4\rangle\\
|2_4\rangle\\
|3_4\rangle\\
|4_4\rangle
\end{array}\right\}.
\end{eqnarray*}

% DIMENSION 6

\section{d=6}

In this dimension it is not known if there exist more than three MUBs.
A possible choice of three is to take the products
\begin{equation*}
\label{dim6}
B_6
=
\left(\begin{array}{cccccc}
1&0&0&0&0&0\\
0&1&0&0&0&0\\
0&0&1&0&0&0\\
0&0&0&1&0&0\\
0&0&0&0&1&0\\
0&0&0&0&0&1
\end{array}\right)
=
\left\{\begin{array}{c}
|0\rangle\\
|1\rangle
\end{array}
\right\}
\otimes
\left\{\begin{array}{c}
|0\rangle\\
|1\rangle\\
|2\rangle
\end{array}
\right\},
\end{equation*}
\begin{equation*}
B_0
=\frac{1}{\sqrt{6}}\left(\begin{array}{cccccc}
1&1&1&1&1&1\\
1&\alpha_3&\alpha_3^2&1&\alpha_3&\alpha_3^2\\
1&\alpha_3^2&\alpha_3&1&\alpha_3^2&\alpha_3\\
1&1&1&-1&-1&-1\\
1&\alpha_3&\alpha_3^2&-1&-\alpha_3&-\alpha_3^2\\
1&\alpha_3^2&\alpha_3&-1&-\alpha_3^2&-\alpha_3
\end{array}\right)
=
\left\{\begin{array}{c}
|0_0\rangle\\
|1_0\rangle
\end{array}
\right\}
\otimes
\left\{\begin{array}{c}
|0_0\rangle\\
|1_0\rangle\\
|2_0\rangle
\end{array}
\right\},
\end{equation*}
\begin{equation*}
B_1
=
\frac{1}{\sqrt{6}}\left(\begin{array}{cccccc}
1&1&1&1&1&1\\
\alpha_3&\alpha_3^2&1&\alpha_3&\alpha_3^2&1\\
\alpha_3&1&\alpha_3^2&\alpha_3&1&\alpha_3^2\\
i&i&i&-i&-i&-i\\
i \alpha_3&i \alpha_3^2&i&-i \alpha_3&-i \alpha_3^2&-i\\
i \alpha_3^2&i&i \alpha_3^2 &-i \alpha_3&-i&-i \alpha_3^2
\end{array}\right)
=
\left\{\begin{array}{l}
|0_1\rangle\\
|1_1\rangle
\end{array}
\right\}
\otimes
\left\{\begin{array}{l}
|0_1\rangle\\
|1_1\rangle\\
|2_1\rangle
\end{array}
\right\},
\end{equation*}
where the kets in two-dimensional vectors refer to qubit MUBs (Appendix A)
and the kets in three-dimensional vectors refer to qutrit MUBs (Appendix B).

% DIMENSION 8

\section{d=8}

The bases of this Appendix present explicitly the result of construction described in section 3.4 of the main text.

\begin{equation*}
B_8 =\frac{1}{2\sqrt{2}}\left(\begin{array}{cccccccc}
1&0&0&0&0&0&0&10\\
0&1&0&0&0&0&0&0\\
0&0&1&0&0&0&0&0\\
0&0&0&1&0&0&0&0\\
0&0&0&0&1&0&0&0\\
0&0&1&0&0&1&0&0\\
0&0&0&0&0&0&1&0\\
0&0&0&0&0&0&0&1
\end{array}\right)=\left\{\begin{array}{c}
|0 0 0\rangle\\
|0 0 1\rangle\\
|0 1 0\rangle\\
|0 1 1\rangle\\
|1 0 0\rangle\\
|1 0 1\rangle\\
|1 1 0\rangle\\
|1 1 1\rangle
\end{array}\right\},
\end{equation*}
\begin{equation*}
B_0=\frac{1}{2\sqrt{2}}\left(\begin{array}{cccccccc}
1&1&1&1&1&1&1&1\\
i&-i&1&-i&i&-i&i&-i\\
i&i&-i&-i&i&i&-i&-i\\
-1&1&1&-1&-1&1&1&-1\\
i&i&i&i&-i&-i&-i&-i\\
-1&1&-1&1&1&-1&1&-1\\
-1&-1&1&1&1&1&-1&-1\\
-i&i&i&-i&i&-i&-i&i
\end{array}\right)=\left\{\begin{array}{c}
|0_00_00_0\rangle\\
|0_00_01_0\rangle\\
|0_01_00_0\rangle\\
|0_01_01_0\rangle\\
|1_00_00_0\rangle\\
|1_00_01_0\rangle\\
|1_01_00_0\rangle\\
|1_01_01_0\rangle
\end{array}\right\},
\end{equation*}
\begin{eqnarray*}
B_1&=&\frac{1}{2\sqrt{2}}\left(\begin{array}{cccccccc}
1&1&1&1&1&1&1&1\\
1&-1&1&-1&1&-1&1&-1\\
i&i&-i&-i&i&i&-i&-i\\
-i&i&i&-i&-i&i&i&-i\\
i&i&i&i&-i&-i&-i&-i\\
i&-i&i&-i&-i&i&-i&i\\
1&1&-1&-1&-1&-1&1&1\\
-1&1&1&-1&1&-1&-1&1
\end{array}\right)
\\
& = &
\frac{1}{\sqrt{2}}\left\{\begin{array}{c}
|0_0 0 0_1\rangle+|1_01 1_1\rangle\\
|0_00 1_1\rangle+|1_01 0_1\rangle\\
|0_01 0_1\rangle+|1_00 1_1\rangle\\
|0_01 1_1\rangle+|1_00 0_1\rangle\\
|0_00 0_1\rangle-|1_01 1_1\rangle\\
|0_00 1_1\rangle-|1_01 0_1\rangle\\
|0_01 0_1\rangle-|1_00 1_1\rangle\\
|0_01 1_1\rangle-|1_00 0_1\rangle\end{array}\right\},
\end{eqnarray*}
\begin{eqnarray*}
B_2&=&\frac{1}{2\sqrt{2}}\left(\begin{array}{cccccccc}
1&1&1&1&1&1&1&1\\
i&-i&i&-i&i&-i&i&-i\\
1&1&-1&-1&1&1&-1&-1\\
-i&i&i&-i&-i&i&i&-i\\
i&i&i&i&-i&-i&-i&-i\\
1&-1&1&-1&-1&1&-1&1\\
-i&-i&i&i&i&i&-i&-i\\
-1&1&1&-1&1&-1&-1&1
\end{array}\right)
\\
& = &
\frac{1}{\sqrt{2}}\left\{\begin{array}{c}
|0 0_10_0\rangle+i|1 1_11_0\rangle\\
|0 0_11_0\rangle+i|1 1_10_0\rangle\\
|0 1_10_0\rangle+i|1 0_11_0\rangle\\
|0 1_11_0\rangle+i|1 0_10_0\rangle\\
|0 0_10_0\rangle-i|1 1_11_0\rangle\\
|0 0_11_0\rangle-i|1 1_10_0\rangle\\
|0 1_10_0\rangle-i|1 0_11_0\rangle\\
|0 1_11_0\rangle-i|1 0_10_0\rangle\end{array}\right\},
\end{eqnarray*}
\begin{eqnarray*}
B_3&=&\frac{1}{2\sqrt{2}}\left(\begin{array}{cccccccc}
1&1&1&1&1&1&1&1\\
1&-1&1&-1&1&-1&1&-1\\
1&1&-1&-1&1&1&-1&-1\\
-1&1&1&-1&-1&1&1&-1\\
i&i&i&i&-i&-i&-i&-i\\
-i&i&-i&i&i&-i&i&-i\\
i&i&-i&-i&-i&-i&i&i\\
i&-i&-i&i&-i&i&i&-i
\end{array}\right)
\\
& = & \frac{1}{\sqrt{2}}\left\{\begin{array}{c}
|0_00_10\rangle+|1_01_11\rangle\\
|0_00_11\rangle+|1_01_10\rangle\\
|0_01_10\rangle+|1_00_11\rangle\\
|0_01_11\rangle+|1_00_10\rangle\\
|0_00_10\rangle-|1_01_11\rangle\\
|0_00_11\rangle-|1_01_10\rangle\\
|0_01_10\rangle-|1_00_11\rangle\\
|0_01_11\rangle-|1_00_10\rangle\end{array}\right\},
\end{eqnarray*}
\begin{eqnarray*}
B_4&=&\frac{1}{2\sqrt{2}}\left(\begin{array}{cccccccc}
1&1&1&1&1&1&1&1\\
i&-i&i&-i&i&-i&i&-i\\
i&i&-i&-i&i&i&-i&-i\\
1&-1&-1&1&1&-1&-1&1\\
1&1&1&1&-1&-1&-1&-1\\
-i&i&-i&i&i&-i&i&-i\\
i&i&-i&-i&-i&-i&i&i\\
-1&1&1&-1&1&-1&-1&1
\end{array}\right)
\\
& = & \left\{\begin{array}{c}
|0_10_00\rangle+i|1_11_01\rangle\\
|0_10_01\rangle+i|1_11_00\rangle\\
|0_11_00\rangle+i|1_10_01\rangle\\
|0_11_01\rangle+i|1_10_00\rangle\\
|0_10_00\rangle-i|1_11_01\rangle\\
|0_10_01\rangle-i|1_11_00\rangle\\
|0_11_00\rangle-i|1_10_01\rangle\\
|0_11_01\rangle-i|1_10_00\rangle\end{array}\right\},
\end{eqnarray*}
\begin{eqnarray*}
B_5&=&\frac{1}{2\sqrt{2}}\left(\begin{array}{cccccccc}
1&1&1&1&1&1&1&1\\
1&-1&1&-1&1&-1&1&-1\\
i&i&-i&-i&i&i&-i&-i\\
i&-i&-i&i&i&-i&-i&i\\
1&1&1&1&-1&-1&-1&-1\\
-1&1&-1&1&1&-1&1&-1\\
-i&-i&i&i&i&i&-i&-i\\
i&-i&-i&i&-i&i&i&-i
\end{array}\right)
\\
& = &
\frac{1}{\sqrt{2}}\left\{\begin{array}{c}
\ket{0 0_00_1}+\ket{1 1_01_1}\\ 
\ket{0 0_01_1}+\ket{1 1_00_1}\\ 
\ket{0 1_00_1}+\ket{1 0_01_1}\\ 
\ket{0 1_01_1}+\ket{1 1_01_1}\\ 
\ket{0 0_00_1}-\ket{1 1_01_1}\\ 
\ket{0 0_01_1}-\ket{1 1_00_1}\\ 
\ket{0 1_00_1}-\ket{1 0_01_1}\\ 
\ket{0 1_01_1}-\ket{1 1_01_1}
\end{array}\right\},
\end{eqnarray*}
\begin{eqnarray*}
B_6&=&\frac{1}{2\sqrt{2}}\left(\begin{array}{cccccccc}
1&1&1&1&1&1&1&1\\
i&-i&i&-i&i&-i&i&-i\\
1&1&-1&-1&1&1&-1&-1\\
-i&i&i&-i&-i&i&i&-i\\
1&1&1&1&-1&-1&-1&-1\\
-i&i&-i&i&i&-i&i&-i\\
-1&-1&1&1&1&1&-1&-1\\
i&-i&-i&i&-i&i&i&-i
\end{array}\right)
\\
& = &
\frac{1}{\sqrt{2}}\left\{\begin{array}{c}
|0 0_00_1\rangle+i|1 1_01_1\rangle\\
|0 0_01_1\rangle+i|1 1_00_1\rangle\\
|0 1_00_1\rangle+i|1 0_01_1\rangle\\
|0 1_01_1\rangle+i|1 0_00_1\rangle\\
|0 0_00_1\rangle-i|1 1_01_1\rangle\\
|0 0_01_1\rangle-i|1 1_00_1\rangle\\
|0 1_00_1\rangle-i|1 0_01_1\rangle\\
|0 1_01_1\rangle-i|1 0_00_1\rangle\end{array}\right\},
\end{eqnarray*}
\begin{equation*}
B_7=\frac{1}{2\sqrt{2}}\left(\begin{array}{cccccccc}
1&1&1&1&1&1&1&1\\
1&-1&1&-1&1&-1&1&-1\\
1&1&-1&-1&1&1&-1&-1\\
1&-1&-1&1&1&-1&-1&1\\
1&1&1&1&-1&-1&-1&-1\\
1&-1&1&-1&-1&1&-1&1\\
1&1&-1&-1&-1&-1&1&1\\
-1&1&1&-1&1&-1&-1&1
\end{array}\right)=\left\{\begin{array}{c}
|0_10_10_1\rangle\\
|0_10_11_1\rangle\\
|0_11_10_1\rangle\\
|0_11_11_1\rangle\\
|1_10_10_1\rangle\\
|1_10_11_1\rangle\\
|1_11_10_1\rangle\\
|1_11_11_1\rangle
\end{array}\right\}
\end{equation*}

% DIMENSION 9

\section{$d=9$}

The bases of this Appendix present explicitly the result of construction described in section 3.3 of the main text for $d=3^2=9$.

\begin{equation*}
B_9
=
\left(\begin{array}{ccccccccc}
1&0&0&0&0&0&0&0&0\\
0&1&0&0&0&0&0&0&0\\
0&0&1&0&0&0&0&0&0\\
0&0&0&1&0&0&0&0&0\\
0&0&0&0&1&0&0&0&0\\
0&0&0&0&0&1&0&0&0\\
0&0&0&0&0&0&1&0&0\\
0&0&0&0&0&0&0&1&0\\
0&0&0&0&0&0&0&0&1
\end{array}\right)
=
\left\{\begin{array}{c}
|0\rangle\\
|1\rangle\\
|2\rangle
\end{array}
\right\}
\otimes
\left\{\begin{array}{c}
|0\rangle\\
|1\rangle\\
|2\rangle
\end{array}
\right\},
\end{equation*}
\begin{equation*}
B_0
=
\frac{1}{3}
\left(\begin{array}{ccccccccc}
1&1&1&1&1&1&1&1&1\\
1&\alpha_3&\alpha_3^2&1&\alpha_3&\alpha_3^2&1&\alpha_3&\alpha_3^2\\
1&\alpha_3^2&\alpha_3&1&\alpha_3^2&\alpha_3&1&\alpha_3^2&\alpha_3\\
1&1&1&\alpha_3&\alpha_3&\alpha_3&\alpha_3^2&\alpha_3^2&\alpha_3^2\\
1&\alpha_3&\alpha_3^2&\alpha_3&\alpha_3^2&1&\alpha_3^2&1&\alpha_3\\
1&\alpha_3^2&\alpha_3&\alpha_3&1&\alpha_3^2&\alpha_3^2&\alpha_3&1\\
1&1&1&\alpha_3^2&\alpha_3^2&\alpha_3^2&\alpha_3&\alpha_3&\alpha_3\\
1&\alpha_3&\alpha_3^2&\alpha_3^2&1&\alpha_3&\alpha_3&\alpha_3^2&1\\
1&\alpha_3^2&\alpha_3&\alpha_3^2&\alpha_3&1&\alpha_3&1&\alpha_3^2
\end{array}\right)
=
\left\{\begin{array}{c}
|0_0\rangle\\
|1_0\rangle\\
|2_0\rangle
\end{array}
\right\}
\otimes
\left\{\begin{array}{c}
|0_0\rangle\\
|1_0\rangle\\
|2_0\rangle
\end{array}
\right\},
\end{equation*}
\begin{equation*}
B_1
=
\frac{1}{3}
\left(\begin{array}{ccccccccc}
1&1&1&1&1&1&1&1&1\\
\alpha_3&\alpha_3^2&1&\alpha_3&\alpha_3^2&1&\alpha_3&\alpha_3^2&1\\
\alpha_3&1&\alpha_3^2&\alpha_3&1&\alpha_3^2&\alpha_3&1&\alpha_3^2\\
\alpha_3&\alpha_3&\alpha_3&\alpha_3^2&\alpha_3^2&\alpha_3^2&1&1&1\\
\alpha_3^2&1&\alpha_3&1&\alpha_3&\alpha_3^2&\alpha_3&\alpha_3^2&1\\
\alpha_3^2&\alpha_3&1&1&\alpha_3^2&\alpha_3&\alpha_3&1&\alpha_3^2\\
\alpha_3&\alpha_3&\alpha_3&1&1&1&\alpha_3^2&\alpha_3^2&\alpha_3^2\\
\alpha_3^2&1&\alpha_3&\alpha_3&\alpha_3^2&1&1&\alpha_3&\alpha_3^2\\
\alpha_3^2&\alpha_3&1&\alpha_3&1&\alpha_3^2&1&\alpha_3^2&\alpha_3
\end{array}\right)
=
\left\{\begin{array}{c}
|0_1\rangle\\
|1_1\rangle\\
|2_1\rangle
\end{array}
\right\}
\otimes
\left\{\begin{array}{c}
|0_1\rangle\\
|1_1\rangle\\
|2_1\rangle
\end{array}
\right\},
\end{equation*}
\begin{equation*}
B_2
=
\frac{1}{3}
\left(\begin{array}{ccccccccc}
1&1&1&1&1&1&1&1&1\\
\alpha_3^2&1&\alpha_3&\alpha_3^2&1&\alpha_3&\alpha_3^2&1&\alpha_3\\
\alpha_3^2&\alpha_3&1&\alpha_3^2&\alpha_3&1&\alpha_3^2&\alpha_3&1\\
\alpha_3^2&\alpha_3^2&\alpha_3^2&1&1&1&\alpha_3&\alpha_3&\alpha_3\\
\alpha_3&\alpha_3^2&1&\alpha_3^2&1&\alpha_3&1&\alpha_3&\alpha_3^2\\
\alpha_3&1&\alpha_3^2&\alpha_3^2&\alpha_3&1&1&\alpha_3^2&\alpha_3\\
\alpha_3^2&\alpha_3^2&\alpha_3^2&\alpha_3&\alpha_3&\alpha_3&1&1&1\\
\alpha_3&\alpha_3^2&1&1&\alpha_3&\alpha_3^2&\alpha_3^2&1&\alpha_3\\
\alpha_3&1&\alpha_3^2&1&\alpha_3^2&\alpha_3&\alpha_3^2&\alpha_3&1
\end{array}\right)
=
\left\{\begin{array}{c}
|0_2\rangle\\
|1_2\rangle\\
|2_2\rangle
\end{array}
\right\}
\otimes
\left\{\begin{array}{c}
|0_2\rangle\\
|1_2\rangle\\
|2_2\rangle
\end{array}
\right\},
\end{equation*}
\begin{eqnarray*}
B_3
& = &
\frac{1}{3}
\left(\begin{array}{ccccccccc}
1&1&1&1&1&1&1&1&1\\
\alpha_3&\alpha_3^2&1&\alpha_3&1&1&\alpha_3&\alpha_3^2&1\\
\alpha_3&1&\alpha_3^2&\alpha_3&\alpha_3^2&\alpha_3^2&\alpha_3&1&\alpha_3^2\\
1&1&1&\alpha_3&1&\alpha_3&\alpha_3^2&\alpha_3^2&\alpha_3^2\\
1&\alpha_3&\alpha_3^2&\alpha_3&\alpha_3^2&1&\alpha_3^2&1&\alpha_3\\
\alpha_3^2&\alpha_3&1&1&\alpha_3^2&\alpha_3&\alpha_3&1&\alpha_3^2\\
1&1&1&\alpha_3^2&\alpha_3^2&\alpha_3^2&\alpha_3&\alpha_3&\alpha_3\\
\alpha_3^2&1&\alpha_3&\alpha_3&\alpha_3^2&1&1&\alpha_3&\alpha_3^2\\
1&\alpha_3^2&\alpha_3&\alpha_3^2&\alpha_3&1&\alpha_3&1&\alpha_3^2
\end{array}\right)
\\
& = &
\frac{1}{\sqrt{3}}
\left\{	
\begin{array}{l}
|0_0 \rangle |0_9 \rangle +\alpha_3 |1_0 \rangle |1_9 \rangle + \alpha_3 |2_0 \rangle | 2_9 \rangle\\
|0_0 \rangle |0_9 \rangle +\alpha_3^2 |1_0 \rangle |1_9 \rangle + |2_0 \rangle | 2_9 \rangle\\
|0_0 \rangle |0_9 \rangle + |1_0 \rangle |1_9 \rangle + \alpha_3^2 |2_0 \rangle | 2_9 \rangle\\
|1_0 \rangle |0_9 \rangle +\alpha_3 |2_0 \rangle |1_9 \rangle + \alpha_3 |0_0 \rangle | 2_9 \rangle \\
|1_0 \rangle |0_9 \rangle +\alpha_3^2 |2_0 \rangle |1_9 \rangle + |0_0 \rangle | 2_9 \rangle\\
|1_0 \rangle |0_9 \rangle + |1_0 \rangle |2_9 \rangle + \alpha_3^2 |0_0 \rangle | 2_9 \rangle \\
|2_0 \rangle |0_9 \rangle +\alpha_3 |0_0 \rangle |1_9 \rangle + \alpha_3 |1_0 \rangle | 2_9 \rangle \\
|2_0 \rangle |0_9 \rangle +\alpha_3^2 |0_0 \rangle |1_9 \rangle + |1_0 \rangle | 2_9 \rangle \\
|2_0 \rangle |0_9 \rangle + |1_0 \rangle |0_9 \rangle + \alpha_3^2 |1_0 \rangle | 2_9 \rangle
\end{array}\right\}
=
\mathcal{P}_3^2
\left[
\left\{\begin{array}{c}
|0_0\rangle\\
|1_0\rangle\\
|2_0\rangle
\end{array}
\right\}
\otimes
\left\{\begin{array}{c}
|0_1\rangle\\
|1_1\rangle\\
|2_1\rangle
\end{array}
\right\}
\right],
\end{eqnarray*}
\begin{eqnarray*}
B_4
& = &
\frac{1}{3}
\left(\begin{array}{ccccccccc}
1&1&1&1&1&1&1&1&1\\
\alpha_3^2&1&\alpha_3&\alpha_3^2&1&\alpha_3&\alpha_3^2&1&\alpha_3\\
\alpha_3^2&\alpha_3&1&\alpha_3^2&\alpha_3&1&\alpha_3^2&\alpha_3&1\\
1&1&1&\alpha_3&\alpha_3&\alpha_3&\alpha_3^2&\alpha_3^2&\alpha_3^2\\
1&\alpha_3&\alpha_3^2&\alpha_3&\alpha_3^2&1&\alpha_3^2&1&\alpha_3\\
\alpha_3&1&\alpha_3^2&\alpha_3^2&\alpha_3&1&1&\alpha_3^2&\alpha_3\\
1&1&1&\alpha_3^2&\alpha_3^2&\alpha_3^2&\alpha_3&\alpha_3&\alpha_3\\
\alpha_3&\alpha_3^2&1&1&\alpha_3&\alpha_3^2&\alpha_3^2&1&\alpha_3\\
1&\alpha_3^2&\alpha_3&\alpha_3^2&\alpha_3&1&\alpha_3&1&\alpha_3^2
\end{array}\right)
\\
& = &
\frac{1}{\sqrt{3}}
\left\{
\begin{array}{l}
|0_0 \rangle |0_9 \rangle +\alpha_3^2 |1_0 \rangle |2_9 \rangle + \alpha_3^2 |2_0 \rangle | 1_9 \rangle\\
|0_0 \rangle |0_9 \rangle +\alpha_3 |1_0 \rangle |2_9 \rangle + |2_0 \rangle | 1_9 \rangle\\
|0_0 \rangle |0_9 \rangle + |1_0 \rangle |2_9 \rangle + \alpha_3 |2_0 \rangle | 1_9 \rangle \\
|1_0 \rangle |0_9 \rangle +\alpha_3^2 |2_0 \rangle |2_9 \rangle + \alpha_3^2 |0_0 \rangle | 1_9 \rangle\\
|1_0 \rangle |0_9 \rangle +\alpha_3 |2_0 \rangle |2_9 \rangle + |0_0 \rangle | 1_9 \rangle \\
|1_0 \rangle |0_9 \rangle + |2_0 \rangle |2_9 \rangle + \alpha_3 |0_0 \rangle | 1_9 \rangle \\
|2_0 \rangle |0_9 \rangle +\alpha_3^2 |0_0 \rangle |2_9 \rangle + \alpha_3^2 |1_0 \rangle | 1_9 \rangle \\
|2_0 \rangle |0_9 \rangle +\alpha_3 |0_0 \rangle |2_9 \rangle + |1_0 \rangle | 1_9 \rangle \\
|2_0 \rangle |0_9 \rangle + |0_0 \rangle |2_9 \rangle + \alpha_3 |1_0 \rangle | 1_9 \rangle
\end{array}\right)
=
\mathcal{P}_3
\left[
\left\{\begin{array}{c}
|0_0\rangle\\
|1_0\rangle\\
|2_0\rangle
\end{array}
\right\}
\otimes
\left\{\begin{array}{c}
|0_2\rangle\\
|1_2\rangle\\
|2_2\rangle
\end{array}
\right\}
\right],
\end{eqnarray*}
\begin{eqnarray*}
B_5
& = &
\frac{1}{3}
\left(\begin{array}{ccccccccc}
1&1&1&1&1&1&1&1&1\\
1&\alpha_3&\alpha_3^2&1&\alpha_3&\alpha_3^2&1&\alpha_3&\alpha_3^2\\
1&\alpha_3^2&\alpha_3&1&\alpha_3^2&\alpha_3&1&\alpha_3^2&\alpha_3\\
\alpha_3&\alpha_3&\alpha_3&\alpha_3^2&\alpha_3^2&\alpha_3^2&1&1&1\\
\alpha_3^2&1&\alpha_3&1&\alpha_3&\alpha_3^2&\alpha_3&\alpha_3^2&1\\
1&\alpha_3^2&\alpha_3&\alpha_3&1&\alpha_3^2&\alpha_3^2&\alpha_3&1\\
\alpha_3&\alpha_3&\alpha_3&1&1&1&\alpha_3^2&\alpha_3^2&\alpha_3^2\\
1&\alpha_3&\alpha_3^2&\alpha_3^2&1&\alpha_3&\alpha_3&\alpha_3^2&1\\
\alpha_3^2&\alpha_3&1&\alpha_3&1&\alpha_3^2&1&\alpha_3^2&\alpha_3
\end{array}\right)
\\
& = &
\frac{1}{\sqrt{3}}
\left\{
\begin{array}{l}
|0_1 \rangle |0_9 \rangle +|1_1 \rangle |2_9 \rangle + |2_1 \rangle | 1_9 \rangle\\
|0_1 \rangle |0_9 \rangle +\alpha_3^2 |1_1 \rangle |2_9 \rangle + \alpha_3 |2_1 \rangle | 1_9 \rangle\\
|0_1 \rangle |0_9 \rangle + \alpha_3 |1_1 \rangle |2_9 \rangle + \alpha_3^2 |2_1 \rangle | 1_9 \rangle\\
|1_1 \rangle |0_9 \rangle +|2_1 \rangle |2_9 \rangle + |0_1 \rangle | 1_9 \rangle\\
|1_1 \rangle |0_9 \rangle +\alpha_3^2 |2_1 \rangle |2_9 \rangle + \alpha_3 |0_1 \rangle | 1_9 \rangle \\
|1_1 \rangle |0_9 \rangle + \alpha_3 |2_1 \rangle |2_9 \rangle + \alpha_3^2 |0_1 \rangle | 1_9 \rangle \\
|2_1 \rangle |0_9 \rangle +|0_1 \rangle |2_9 \rangle + |1_1 \rangle | 1_9 \rangle \\
|2_1 \rangle |0_9 \rangle +\alpha_3^2 |0_1 \rangle |2_9 \rangle + \alpha_3 |1_1 \rangle | 1_9 \rangle\\
|2_1 \rangle |0_9 \rangle + \alpha_3 |0_1 \rangle |2_9 \rangle + \alpha_3^2 |1_1 \rangle | 1_9 \rangle
\end{array}\right\}
=
\mathcal{P}_3
\left[
\left\{\begin{array}{c}
|0_1\rangle\\
|1_1\rangle\\
|2_1\rangle
\end{array}
\right\}
\otimes
\left\{\begin{array}{c}
|0_0\rangle\\
|1_0\rangle\\
|2_0\rangle
\end{array}
\right\}
\right],
\end{eqnarray*}
\begin{eqnarray*}
B_6
& = &
\frac{1}{3}
\left(\begin{array}{ccccccccc}
1&1&1&1&1&1&1&1&1\\
\alpha_3^2&1&\alpha_3&\alpha_3^2&1&\alpha_3&\alpha_3^2&1&\alpha_3\\
\alpha_3^2&\alpha_3&1&\alpha_3^2&\alpha_3&1&\alpha_3^2&\alpha_3&1\\
\alpha_3&\alpha_3&\alpha_3&\alpha_3^2&\alpha_3^2&\alpha_3^2&1&1&1\\
\alpha_3^2&1&\alpha_3&1&\alpha_3&\alpha_3^2&\alpha_3&\alpha_3^2&1\\
\alpha_3&1&\alpha_3^2&\alpha_3^2&\alpha_3&1&1&\alpha_3^2&\alpha_3\\
\alpha_3&\alpha_3&\alpha_3&1&1&1&\alpha_3^2&\alpha_3^2&\alpha_3^2\\
\alpha_3&\alpha_3^2&1&1&\alpha_3&\alpha_3^2&\alpha_3^2&1&\alpha_3\\
\alpha_3^2&\alpha_3&1&\alpha_3&1&\alpha_3^2&1&\alpha_3^2&\alpha_3
\end{array}\right)
\\
& = &
\frac{1}{\sqrt{3}}
\left\{
\begin{array}{l}
|0_1 \rangle |0_9 \rangle + \alpha_3^2 |1_1 \rangle |1_9 \rangle + \alpha_3^2 |2_1 \rangle | 2_9 \rangle\\
|0_1 \rangle |0_9 \rangle + |1_1 \rangle |1_9 \rangle + \alpha_3 |2_1 \rangle | 2_9 \rangle \\
|0_1 \rangle |0_9 \rangle + \alpha_3 |1_1 \rangle |1_9 \rangle +|2_1 \rangle |2_9 \rangle\\
|1_1 \rangle |0_9 \rangle + \alpha_3^2 |2_1 \rangle |1_9 \rangle + \alpha_3^2 |0_1 \rangle | 2_9 \rangle\\
|1_1 \rangle |0_9 \rangle + |2_1 \rangle |1_9 \rangle + \alpha_3 |0_1 \rangle | 2_9 \rangle \\
|1_1 \rangle |0_9 \rangle + \alpha_3 |2_1 \rangle |1_9 \rangle +|0_1 \rangle |2_9 \rangle \\
|2_1 \rangle |0_9 \rangle + \alpha_3^2 |0_1 \rangle |1_9 \rangle + \alpha_3^2 |1_1 \rangle | 2_9 \rangle \\
|2_1 \rangle |0_9 \rangle + |0_1 \rangle |1_9 \rangle + \alpha_3 |1_1 \rangle | 2_9 \rangle \\
|2_1 \rangle |0_9 \rangle + \alpha_3 |0_1 \rangle |1_9 \rangle +|1_1 \rangle |2_9 \rangle
\end{array}\right\}
=
\mathcal{P}_3^2
\left[
\left\{\begin{array}{c}
|0_1\rangle\\
|1_1\rangle\\
|2_1\rangle
\end{array}
\right\}
\otimes
\left\{\begin{array}{c}
|0_2\rangle\\
|1_2\rangle\\
|2_2\rangle
\end{array}
\right\}
\right],
\end{eqnarray*}
\begin{eqnarray*}
B_7
& = &
\frac{1}{3}
\left(\begin{array}{ccccccccc}
1&1&1&1&1&1&1&1&1\\
1&\alpha_3&\alpha_3^2&1&\alpha_3&\alpha_3^2&1&\alpha_3&\alpha_3^2\\
1&\alpha_3^2&\alpha_3&1&\alpha_3^2&\alpha_3&1&\alpha_3^2&\alpha_3\\
\alpha_3^2&\alpha_3^2&\alpha_3^2&1&1&1&\alpha_3&\alpha_3&\alpha_3\\
\alpha_3&\alpha_3^2&1&\alpha_3^2&1&\alpha_3&1&\alpha_3&\alpha_3^2\\
1&\alpha_3^2&\alpha_3&\alpha_3&1&\alpha_3^2&\alpha_3^2&\alpha_3&1\\
\alpha_3^2&\alpha_3^2&\alpha_3^2&\alpha_3&\alpha_3&\alpha_3&1&1&1\\
1&\alpha_3&\alpha_3^2&\alpha_3^2&1&\alpha_3&\alpha_3&\alpha_3^2&1\\
\alpha_3&1&\alpha_3^2&1&\alpha_3^2&\alpha_3&\alpha_3^2&\alpha_3&1
\end{array}\right)
\\
& = &
\frac{1}{\sqrt{3}}
\left\{
\begin{array}{l}
|0_2 \rangle |0_9 \rangle + |1_2 \rangle |1_9 \rangle + |2_2 \rangle | 2_9 \rangle \\
|0_2 \rangle |0_9 \rangle + \alpha_3 |1_2 \rangle |1_9 \rangle + \alpha_3^2 |2_2 \rangle | 2_9 \rangle \\
|0_2 \rangle |0_9 \rangle + \alpha_3^2 |1_2 \rangle |1_9 \rangle + \alpha_3 |2_2 \rangle |2_9 \rangle \\
|1_2 \rangle |0_9 \rangle + |2_2 \rangle |1_9 \rangle + |0_2 \rangle | 2_9 \rangle\\
|1_2 \rangle |0_9 \rangle + \alpha_3 |2_2 \rangle |1_9 \rangle + \alpha_3^2 |0_2 \rangle | 2_9 \rangle \\
|1_2 \rangle |0_9 \rangle + \alpha_3^2 |2_2 \rangle |1_9 \rangle + \alpha_3 |0_2 \rangle |2_9 \rangle \\
|2_2 \rangle |0_9 \rangle + |0_2 \rangle |1_9 \rangle + |1_2 \rangle | 2_9 \rangle\\
|2_2 \rangle |0_9 \rangle + \alpha_3 |0_2 \rangle |1_9 \rangle + \alpha_3^2 |1_2 \rangle | 2_9 \rangle\\
|2_2 \rangle |0_9 \rangle + \alpha_3^2 |0_2 \rangle |1_9 \rangle + \alpha_3 |1_2 \rangle |2_9  \rangle
\end{array}\right\}
=
\mathcal{P}_3^2
\left[
\left\{\begin{array}{c}
|0_2\rangle\\
|1_2\rangle\\
|2_2\rangle
\end{array}
\right\}
\otimes
\left\{\begin{array}{c}
|0_0\rangle\\
|1_0\rangle\\
|2_0\rangle
\end{array}
\right\}
\right],
\end{eqnarray*}
\begin{eqnarray*}
B_8
& = &
\frac{1}{3}
\left(\begin{array}{ccccccccc}
1&1&1&1&1&1&1&1&1\\
\alpha_3&\alpha_3^2&1&\alpha_3&\alpha_3^2&1&\alpha_3&\alpha_3^2&1\\
\alpha_3&1&\alpha_3^2&\alpha_3&1&\alpha_3^2&\alpha_3&1&\alpha_3^2\\
\alpha_3^2&\alpha_3^2&\alpha_3^2&1&1&1&\alpha_3&\alpha_3&\alpha_3\\
\alpha_3&\alpha_3^2&1&\alpha_3^2&1&\alpha_3&1&\alpha_3&\alpha_3^2\\
\alpha_3^2&\alpha_3&1&1&\alpha_3^2&\alpha_3&\alpha_3&1&\alpha_3^2\\
\alpha_3^2&\alpha_3^2&\alpha_3^2&\alpha_3&\alpha_3&\alpha_3&1&1&1\\
\alpha_3^2&1&\alpha_3&\alpha_3&\alpha_3^2&1&1&\alpha_3&\alpha_3^2\\
\alpha_3&1&\alpha_3^2&1&\alpha_3^2&\alpha_3&\alpha_3^2&\alpha_3&1
\end{array}\right)
\\
& = &
\frac{1}{\sqrt{3}}
\left\{
\begin{array}{l}
|0_2 \rangle |0_9 \rangle + \alpha_3 |1_2 \rangle |2_9 \rangle + \alpha_3 |2_2 \rangle | 1_9 \rangle \\
|0_2 \rangle |0_9 \rangle + |1_2 \rangle |2_9 \rangle + \alpha_3^2 |2_2 \rangle |1_9 \rangle \\
|0_2 \rangle |0_9 \rangle + \alpha_3^2 |1_2 \rangle |2_9 \rangle + |2_2 \rangle |1_9 \rangle \\
|1_2 \rangle |0_9 \rangle + \alpha_3 |2_2 \rangle |2_9 \rangle + \alpha_3 |0_2 \rangle | 1_9 \rangle \\
|1_2 \rangle |0_9 \rangle + |2_2 \rangle |2_9 \rangle + \alpha_3^2 |0_2 \rangle |1_9 \rangle \\
|1_2 \rangle |0_9 \rangle + \alpha_3^2 |2_2 \rangle |2_9 \rangle + |0_2 \rangle |1_9 \rangle\\
|2_2 \rangle |0_9 \rangle + \alpha_3 |0_2 \rangle |2_9 \rangle + \alpha_3 |1_2 \rangle | 1_9 \rangle \\
|2_2 \rangle |0_9 \rangle + |0_2 \rangle |2_9 \rangle + \alpha_3^2 |1_2 \rangle |1_9 \rangle \\
|2_2 \rangle |0_9 \rangle + \alpha_3^2 |0_2 \rangle |2_9 \rangle + |1_2 \rangle |1_9 \rangle
\end{array}\right\}
=
\mathcal{P}_3
\left[
\left\{\begin{array}{c}
|0_2\rangle\\
|1_2\rangle\\
|2_2\rangle
\end{array}
\right\}
\otimes
\left\{\begin{array}{c}
|0_1\rangle\\
|1_1\rangle\\
|2_1\rangle
\end{array}
\right\}
\right],\end{eqnarray*}
where the kets refer to MUBs for a qutrit (Appendix B),
and $\mathcal{P}_3$ is the control-phase gate for two qutrits
as given in Eq. (\ref{controlphase}) of the main text.

\end{document}